\documentclass[preprint,showpacs,preprintnumbers,amsmath,amssymb]{revtex4}

\usepackage{graphicx}
\usepackage{dcolumn}
\usepackage{bm}

\begin{document}

\preprint{APS/123-QED}

\title{New Structural Quantum Circuit Simulating a Toffoli Gate}

\author{Masanari Asano}

 \email{asano@is.noda.tus.ac.jp}
\affiliation{%
Dept.\ of Information Sciences, 
  Faculty of Science \& Technology,
  Tokyo University of Science,
  Noda City, Chiba 278, Japan.}%

\author{Chikara Ishii}%
\affiliation{%
Dept.\ of Physics, Faculty of Science, Tokyo University of Science, 1-3 Kagurazaka, Shinjuku-ku, Tokyo 162-8601, Japan.
}%

\date{\today}

\begin{abstract}
A Toffoli gate ($C^{n}$-NOT gate) is regarded as an important unitary gate in quantum computation, and is simulated by a quantum circuit composed of $C^{2}$-NOT gates. This paper presents a quantum circuit with a new configuration of $C^{2}$-NOT gates simulating a $C^{2m+1}$-NOT operation under the condition $m=2^{n}$ ($n=1,2,\cdots$). The circuit is composed of units called multi-qubits gates (MQGs), each of which performs $m$ $C^{2}$-NOT operations simultaneously on $3m$ qubits. Simultaneous operations eliminate the need to manipulate qubits individually, as required in conventional quantum circuits. The proposed circuit thus represents a more realistic mode of operation for practical computing systems. A nuclear magnetic resonance implementation of the circuit is presented as a demonstration of the feasibility of MQG operations for practical systems.
\end{abstract}

\pacs{03.67.Lx}
                             
\keywords{quantum computation, quantum circuit, multiple qubit CNOT gate}
\maketitle

\section{Introduction}

An ideal quantum computer is a physical architecture that implements any unitary operation $[U(2^{n})]$ on $n$ qubits, and quantum computation (QC) is expressed by a quantum circuit composed of universal gates, such as one-bit gates $[U(2)]$ and controlled-NOT gates~\cite{gate}. Simple QC on a few qubits has already been demonstrated experimentally in various systems, including nuclear magnetic resonance (NMR)~\cite{NMR}, ion-trap~\cite{iontrap}, superconducting qubit~\cite{scq}, and all-optical~\cite{optical} systems. More recently, several architectures consisting of many particles as two-level quantum systems have also been proposed for larger-scale computations~\cite{silicon,S-NMR1,S-NMR2}. However, it remains difficult with the present technical capabilities for the manipulation of qubits to realize large-scale unitary operations over quantum circuits, even if a system with many qubits can be successfully fabricated. Quantum circuits are usually designed on the assumption that qubits can be manipulate individually, requiring the skillful application of some external effect to cause a subtle change in the state of a single particle. Furthermore, control of quantum correlations between particles is necessary to realize gate operations, similarly requiring skillful control of switch interactions between qubits. The magnitudes of interactions are often restricted in practical systems, and are dependent on the configuration of particles. Additional operations (e.g., swap operations) are often needed to realize gate operations for qubits without interaction. As the number of qubits increases, such tasks become too delicate and too complicated to be accomplished within the coherence time of physical systems. \par
The present study introduces a quantum circuit designed to avoid the above technical problems regarding the manipulation of qubits in large-scale computation. The circuit proposed here simulates a specific multiple-qubit gate -- the Toffoli gate ($C^{N-1}$-NOT gate)~\cite{univ}, which is regarded as an important unit for constructing $2^{N}$-dimensional unitary systems $[U(2^{N})]$~\cite{QC}. This gate plays a critical role in Grover's search algorithm~\cite{grover1,grover2}.  The proposed circuit performs $m$ $C^{2}$-NOT operations simultaneously on $3m$ qubits per process, where the set of $m$ $C^{2}$-NOT gates is referred to as a multi-qubit gate (MQG). It is clarified in this paper that $2m$ MQG operations simulate $C^{2m+1}$-NOT operations under the condition $m=2^{n}$ ($n=1,2,\cdots$). This configuration of $C^{2}$-NOT gates differs considerably from conventional circuits, in which only one $C^{2}$-NOT is performed per process. \par
The $3m$ qubits included in an MQG each consist of two control bits and one target bit, and $m$ qubits of the same type are given common manipulations to realize the operation. Such manipulations can be performed by application of an external effect that causes the states of all these qubits to change at the same time, eliminating the need to treat qubits individually. Another advantage of the new method is that the simple configuration of $C^{2}$-NOT gates in an MQG makes it possible to realize operations in practical systems, where the interactions of qubits are restricted. Accordingly, the technical tasks for manipulating qubits are simplified and reduced in number, regardless of the number of qubits.\par
The remainder of this paper is organized as follows. In Sec.~II, the new quantum circuit simulating a Toffoli gate is introduced, and the MQG structure is presented. The proposed circuit is used to simulate $C^{5}$-NOT operations as an example of the implementation. In Sec.~III, an NMR implementation of the MQG operation is briefly discussed, assuming a triangular lattice as a simple physical system with many qubits. The feasibility of the circuit is demonstrated by showing the manipulation of qubits using resonance fields. The study is finally summarized in Sec.~IV.

\section{Quantum Circuit Simulating a $C^{2^{n+1}+1}$-NOT Gate}

Figure 1 shows a quantum circuit that simulates a $C^{2^{n+1}+1}$-NOT gate. In this section, the characteristics of the circuit are described, and the process of changing the states of qubits is explained through an example implementation. 

\subsection{Circuit Characteristics }
The proposed quantum circuit (Fig. 1) includes $2^{n+2}+1$ qubits, where the $2^{n+1}+1$ bits labeled $a_{0}$,$b_{i}$ and $c_{i}$ ($i=1,2, \cdots ,2^{n}$) act as control bits, and the ($2^{n+2}+1$)-th bit labeled $a_{2^{n}}$ is the target. Here, $A$, $B$, $C$ and $D$ ($A'$, $B'$, $C'$ and $D'$) $\in \{ 0,1 \}$ denote the input (output) basis for the qubits labeled $a$, $b$, $c$ and $d$. As proven in the appendix, the outputs $A'$, $B'$, $C'$ and $D'$ are given by
\begin{eqnarray}
&&A'_{l=2^{n}}=A_{0}\wedge (B_{1}\wedge C_{1})\wedge \cdots (B_{2^{n}}\wedge C_{2^{n}})\oplus A_{l=2^{n}}\nonumber\\
&&A'_{l<2^{n}}=A_{l< 2^{n}},\ B'_{l}=B_{l},\ C'_{l}=C_{l},\ D'_{l}=D_{l}. 
\end{eqnarray}
The solid frame in Fig. 1 contains $2^{n}$ $C^{2}$-NOT gates that transform the states of $2^{n}$ qubits labeled $d$ simultaneously. These gates are regarded as a unit gate, that is, an MQG. Table 1 shows a comparison of the total numbers of units in the proposed circuit and a conventional circuit in which $C^{2}$-NOT operations are performed individually. \par
The number of qubits in the proposed circuit is restricted to $2^{n+2}+1$, but the high expected functionality of the circuit outweighs this restriction. Most importantly, in the operation of the MQG, all qubits with the same label undergo the same transformation, that is, identically labeled qubits can be manipulated as a single qubit. This means that the skill and time required for the manipulation of $3\cdot 2^{n}$ qubits are similar to those required for a $C^{2}$-NOT gate on $3$ qubits, regardless of the number of qubits. Such MQG operations can be readily realized in practical systems without strong long-distance interactions, representing another advantage attributable to the simplicity of the circuit. 

\begin{table*}[htbp]
 \begin{center}
  \begin{tabular}{|c|c|c|c|c|}
    \hline
       & Unit   & Number of units   &  Simulated gate &  Number of qubits   \\
    \hline
     Conventional method  & C$^{2}$-NOT   &  $4(N-3)$  &  C$^{N-1}$-NOT  & $N>6$   \\
    \hline
     Proposed method  &  MQG  & $2N-4$   &  C$^{N-1}$-NOT  &   $N=2^{n+1}+2$ $(n=1,2,\cdots)$ \\
    \hline
  \end{tabular}
 \end{center}
 \caption{Comparison of unit numbers in the proposed circuits and conventional circuits}
\end{table*}

\subsection{ Simulation of $C^{5}$-NOT}

In the quantum circuit composed of MQGs, the state of the input is changed by some complex process. Here, the transformations in a quantum circuit simulating a $C^{5}$-NOT gate are shown as an example of the operation of an MQG circuit (see Fig.~2). The bases of inputs $A$ and $D$ $\in \{0,1\}$ are transformed to $A'$ and $D'$. This circuit is divided into a number of blocks $BLOCK_{l}(k)$ (dotted frames), where the bases of the input/output in each block are defined as $B_{l}$, $C_{l}$, $D_{l}(k-1)$ and $A_{l}(k-1)$ / $B_{l}$, $C_{l}$, $D_{l}(k)$ and $A_{l}(k)$ (see Fig.~3). $A_{l}(k-1)$ is transformed to $A_{l}(k)$ through $Z_{l}(k)$. It can be readily checked that the bases $A_{l}(k)$ and $Z_{l}(k)$ satisfy the following equation:
\begin{eqnarray}
A_{l}(k)=B_{l}\wedge C_{l}\wedge Z_{l-1}(k)\oplus A_{l}(k-1),\label{eq:1}\\
Z_{l}(k)=B_{l}\wedge C_{l}\wedge A_{l-1}(k-1)\oplus Z_{l}(k-1),\label{eq:2}
\end{eqnarray}
where $A_{0}(k)=A_{0}$, $A_{l}(0)=A_{l}$, $Z_{0}(k)=A_{0}$ and $Z_{l}(1)=B_{l}\wedge (A_{l-1}\wedge C_{l}\oplus D_{l})\oplus A_{l}$. From Eqs.~(\ref{eq:1} and \ref{eq:2}), $A'_{1}$ and $A'_{2}$ can be expressed as 
\begin{eqnarray}
A'_{1}&=&A_{1}(2)\nonumber\\
&=&B_{1}\wedge C_{1} \wedge Z_{0}(2) \oplus A_{1}(1)\nonumber\\
&=&B_{1}\wedge C_{1} \wedge Z_{0}(2) \oplus B_{1}\wedge C_{1} \wedge Z_{0}(1)\oplus A_{1}(0)\nonumber\\
&=&A_{1},\label{eq:3}\\
A'_{2}&=&A_{2}(2)\nonumber\\
&=&B_{2}\wedge C_{2}\wedge Z_{1}(2)\oplus A_{2}(1)\nonumber\\
&=&B_{2}\wedge C_{2} \wedge (B_{1}\wedge C_{1}\wedge A_{0}\oplus Z_{1}(1))\nonumber\\
&&\oplus B_{2}\wedge C_{2}\wedge Z_{1}(1)\oplus A_{2}\nonumber\\
&=&A_{0}\wedge B_{1}\wedge C_{1}\wedge B_{2}\wedge C_{2}\oplus A_{2}.\label{eq:4}
\end{eqnarray}

Then, noting the equations,
\begin{eqnarray}
Z_{1}(2)&=&B_{1}\wedge D_{1}\oplus A_{1},\label{eq:5}\\
Z_{2}(2)&=&A_{0}\wedge B_{1}\wedge C_{1}\wedge B_{2}\wedge C_{2}\oplus B_{2}\wedge D_{2}\nonumber\\
&&\oplus A_{2},\label{eq:6}
\end{eqnarray}
and the relation of $B_{l}\wedge D'_{l}\oplus Z_{l}(2)=A'_{l}$, we have the result $D'_{l}=D_{l}$.

\section{MQG Operation Using NMR }
In this section, a method for tuning the interactions required for realizing MQG operations is examined.

\subsection{Implementation of a Multi-qubit System}
As a physical system containing multiple qubits, a simple structural model of a triangular lattice is considered (Fig. 4). Here, $A$, $B$, $C$ and $D$ represent nuclear spins ($S=1/2$), which realize quantum bits corresponding to the labels $a$, $b$, $c$ and $d$ in the quantum circuit shown in Fig. 1. The Hamiltonian of this model is defined as 
\begin{eqnarray}
H&=&\sum_{l=1}^{N/4}aZ_{Al}\cdot Z_{Cl}+bZ_{Cl}\cdot Z_{Dl}+cZ_{Dl}\cdot Z_{Al}\nonumber\\
&&+dZ_{Dl}\cdot Z_{Bl}+eZ_{Bl}\cdot Z_{Al+1}\nonumber\\&&+fZ_{Al+1}\cdot Z_{Dl}.\label{eq:7}
\end{eqnarray}
The terms $Z_{i}Z_{j}$ ($Z$ is a Pauli operator) in the above equation represent nearest-neighbor interactions connecting qubits (solid and dotted lines in Fig. 4), the magnitudes of which are given by $a,b,\ldots,f$. It is realistic that nearest-neighbor interactions are dominant in physical systems, and it can be expected that such a simple structural system may be fabricated artificially. 

\subsection{Tuning Interactions for MQG Operation}
The manipulation of qubits in an NMR quantum computer is performed by applying resonance magnetic fields at appropriate Larmor frequencies, the values of which depend on the physical states of spins with respect to the kinds of atoms and chemical environment. To access all $N$ spins in a system individually, the states must be designed such that each spin can be modified via a different Larmor frequency. When $N$ is large, these differences are very small, and it is inevitable that the manipulation of qubits becomes sensitive and complex in large-scale systems. \par
The method proposed in this study avoids these difficulties by eliminating the need to manipulate qubits individually. This is particularly advantageous when the number of qubits is large. MQG operation can always be realized by applying six cases of resonance field.
The resonance fields are set according to Lamor frequencies of $\omega_{A}$, $\omega_{A'}$, $\omega_{B}$, $\omega_{C}$, $\omega_{D}$ and $\omega_{D'}$, corresponding to qubits  $A_{2n-1}$, $A_{2n}$, $B_{l}$, $C_{l}$, $D_{2n-1}$ and $D_{2n}$ in Fig.~4. It should be noted that six different physical states are given to each spin according to these labels. The preparation of such conditions can be achieved much more readily than the case of $N$ different state for all $N$ qubits. \par
An MQG is a group of $C^{2}$-NOT gates. The spins of $A_{l}$ and $C_{l}$ (or $B_{l}$ and $D_{l}$) correspond to control bits, and $D_{l}$ (or $ A_{l+1} $) acts as a target. To realize the operations of $C^{2}$-NOT operations simultaneously, the interactions denoted $A_{l}-C_{l}$, $C_{l}-D_{l}$ and $D_{l}-A_{l}$ (solid lines in Fig.4) or $D_{l}-B_{l}$, $B_{l}-A_{l+1}$ and $A_{l+1}-D_{l}$ (dotted lines) are employed. The effects of these interactions can be extracted from the temporal evolution of the system by applying the above six kinds of resonance fields to the system instantaneously, as expressed by

\begin{eqnarray}
&&e^{-iHt}R_{D}R_{D'}e^{-iHt}R_{D}R_{D'}R_{B}e^{-iHt}\nonumber\\ &&\times R_{D}R_{D'}e^{-iHt}R_{D}R_{D'}R_{B}=e^{-i4t\sum_{l=1}^{N/4}aZ_{Al}\cdot Z_{Cl}},\nonumber\\
&&e^{-iHt}R_{A}R_{A'}e^{-iHt}R_{A}R_{A'}R_{B}e^{-iHt}\nonumber\\&& \times R_{A}R_{A'}e^{-iHt}R_{A}R_{A'}R_{B}=e^{-i4t\sum_{l=1}^{N/4}bZ_{Cl}\cdot Z_{Dl}},\nonumber\\
&&e^{-iHt}R_{B}R_{C}e^{-iHt}R_{B}R_{C}R_{A'}R_{D'} e^{-iHt}\nonumber\\&& \times R_{B}R_{C}e^{-iHt}R_{B}R_{C}R_{A'}R_{D'}=e^{-i4t\sum_{l=1}^{N/4}cZ_{Dl}\cdot Z_{Al}},\nonumber\\
&&e^{-iHt}R_{A}R_{A'}e^{-iHt}R_{A}R_{A'}R_{C}e^{-iHt}\\&& \times R_{A}R_{A'}e^{-iHt}R_{A}R_{A'}R_{C}=e^{-i4t\sum_{l=1}^{N/4}dZ_{Dl}\cdot Z_{Bl}},\nonumber\\
&&e^{-iHt}R_{D}R_{D'}e^{-iHt}R_{D}R_{D'}R_{C}e^{-iHt}\nonumber\\&& \times R_{D}R_{D'}e^{-iHt}R_{D}R_{D'}R_{C}=e^{-i4t\sum_{l=1}^{N/4}eZ_{Bl}\cdot Z_{Al+1}},\nonumber\\
&&e^{-iHt}R_{B}R_{C}e^{-iHt}R_{B}R_{C}R_{A}R_{D}e^{-iHt}\nonumber\\&& \times R_{B}R_{C}e^{-iHt}R_{B}R_{C}R_{A}R_{D}=e^{-i4t\sum_{l=1}^{N/4}fZ_{Al+1}\cdot Z_{Dl}}.\nonumber
\end{eqnarray}

In these equations, the operators $R_{K}$($K=A,A',B,C,D,D'$) represent the effects of the resonance fields that rotate spins by $180^{\circ}$ about the $x$ axis. The application of a single field with frequency $\omega_{K}$ causes all spins with label $K$ to undergo a simultaneous transformation of rotation. $R_{K}$ are defined as
\begin{eqnarray}
&&R_{A}=e^{-i\frac{\pi}{2} \sum_{n=1}^{N/8} X_{A2n-1}},\nonumber\\
&&R_{A'}=e^{-i\frac{\pi}{2} \sum_{n=1}^{N/8} X_{A2n}},\nonumber\\
&&R_{B}=e^{-i\frac{\pi}{2} \sum_{l=1}^{N/4} X_{Bl}},\\
&&R_{C}=e^{-i\frac{\pi}{2} \sum_{l=1}^{N/4} X_{Cl}},\nonumber\\
&&R_{D}=e^{-i\frac{\pi}{2} \sum_{n=1}^{N/8} X_{D2n-1}},\nonumber\\
&&R_{D'}=e^{-i\frac{\pi}{2} \sum_{n=1}^{N/8} X_{D2n}},\nonumber
\end{eqnarray}
where $X$ is the Pauli operator.

\section{Conclusion}
A new construction of a Toffoli gate ($C^{N-1}$-NOT gate) was presented. The process is expressed as a quantum circuit with simple configuration of multi-qubit gates (MQGs), each of which implements $2^{n}$ $C^{2}$-NOT operations simultaneously. As $N$ in $C^{N-1}$-NOT is restricted to $2^{n+1}+2$ $(n=1,2,\cdots)$ in the proposed system, the operation of $C^{N'-1}$-NOT ($N'\neq 2^{n+1}+2$) is implemented by fixing the states of unused qubits to $\left| 1 \right>$. However, while this circuit is restricted in the number of qubits it must contain, it offers much greater feasibility for practical implementation. For example, MQG operation can be realized using realistic simply structured systems, and does not involve the individual manipulation of qubits.  These advantages are expected to be valuable in a range of systems, including the example NMR scheme briefly discussed in this paper. This study has shown that the technical practicality of operations in quantum circuits can be incorporated into quantum circuit design without loss of functionality.

\appendix

\section{Proof of Equation~(1)}

Equation.~(1) expresses the relation between inputs and outputs in the proposed circuit. This result is derived as follows.\par 

The quantum circuit is divided into blocks $BLOCK_{l}(k)$ ($l,k=1,\cdots,2^{n}$) as expressed in Fig. 3, in which the bases of the input/output satisfy 
\begin{eqnarray}
&&A_{l}(k)=B_{l}\wedge C_{l}\wedge Z_{l-1}(k)\oplus A_{l}(k-1),\label{eq:A1}\\
&&Z_{l}(k)=B_{l}\wedge C_{l}\wedge A_{l-1}(k-1)\oplus Z_{l}(k-1),\label{eq:A2}\\
&&A_{0}(k)=Z_{0}(k)=A_{0}.\label{eq:A3}
\end{eqnarray}
Using Eqs.~(\ref{eq:A1}) and (\ref{eq:A2}), we then have
\begin{eqnarray}
A_{l}(k)&=&B_{l}\wedge C_{l}\wedge \big(B_{l-1}\wedge C_{l-1}\wedge A_{l-2} (k-1)\oplus Z_{l-1}(k-1)\big)\nonumber\\
&&\ \ \ \ \  \oplus \big(B_{l}\wedge C_{l}\wedge Z_{l-1} (k-1)\oplus A_{l}(k-2)\big)\nonumber\\
&=&\big[\bigwedge^{1}_{p=0}(B_{l-p}\wedge C_{l-p})\big] \wedge A_{l-2}(k-1)\oplus A_{l}(k-2).\label{eq:A4}
\end{eqnarray}
Here, $\bigwedge^{m}_{p=0}(B_{l-p}\wedge C_{l-p})$ represents $(B_{l}\wedge C_{l})\wedge (B_{l-1}\wedge C_{l-1})\wedge \cdots \wedge (B_{l-m}\wedge C_{l-m})$. The bases $A_{l-2}(k-1)$ and $A_{l}(k-2)$ in the above equation are similarly represented as
\begin{eqnarray*}
A_{l-2}(k-1)&=&\big[\bigwedge_{p=2}^{3}(B_{l-p}\wedge C_{l-p})\big]\wedge A_{l-4}(k-2)\oplus A_{l-2}(k-3),\\
A_{l}(k-2)&=&\big[\bigwedge_{p=0}^{1}(B_{l-p}\wedge C_{l-p})\big]\wedge A_{l-2}(k-3)\oplus A_{l}(k-4).
\end{eqnarray*}
Therefore, Eq.~(\ref{eq:A4}) is rewritten as
\begin{eqnarray}
A_{l}(k)=\big[\bigwedge_{p=0}^{3}(B_{l-p}\wedge C_{l-p})\big]\wedge A_{l-4}(k-2)\oplus A_{l}(k-4).\label{eq:A4-2}
\end{eqnarray}
Representing the form of $A_{l}(k)$ in this way gives the following general form inductively:

\begin{eqnarray} 
A_{l}(k)=\big[\bigwedge_{p=0}^{2^{n}-1}(B_{l-p}\wedge C_{l-p})\big]\wedge A_{l-2^{n}}(k-2^{n-1})\oplus A_{l}(k-2^{n})\label{eq:A5}.
\end{eqnarray}
In the case of $k=2^{n}$ and $l=2^{n}$, this equation becomes
\begin{equation}
A_{2^{n}}(2^{n})=\big[\bigwedge_{p=0}^{2^{n}-1}(B_{2^{n}-p}\wedge C_{2^{n}-p})\big]\wedge A_{0}(2^{n-1})\oplus A_{2^{n}}(0).
\end{equation}
From $A_{2^{n}}(2^{n})=A'_{2^{n}}$, $A_{0}(2^{n-1})=A_{0}$, and $A_{2^{n}}(0)=A_{2^{n}}$, the relation $A'_{l=2^{n}}=A_{0}\wedge (B_{1}\wedge C_{1})\wedge \cdots (B_{2^{n}}\wedge C_{2^{n}})\oplus A_{l=2^{n}}$ in Eq.~(1) is proven.\par 

To derive the relation $A'_{l< 2^{n}}=A_{l< 2^{n}}(2^{n})=A_{l< 2^{n}}(0)$, we consider the case of $k=2^{n}$ and $l=2^{m}$ ($m<n$), for which Eq.~(\ref{eq:A5}) is given by
\begin{eqnarray}
A_{2^{m}}(2^{n})&=&\big[\bigwedge_{p=0}^{2^{m}-1}(B_{2^{m}-p}\wedge C_{2^{m}-p})\big]\wedge A_{0}(2^{n}-2^{m-1})\oplus A_{2^{m}}(2^{n}-2^{m})\nonumber\\
&=&\big[\bigwedge_{p=0}^{2^{m}-1}(B_{2^{m}-p}\wedge C_{2^{m}-p})\big]\wedge A_{0}\oplus \big[\bigwedge_{p=0}^{2^{m}-1}(B_{2^{m}-p}\wedge C_{2^{m}-p})\big]\wedge A_{0}\oplus A_{2^{m}}(2^{n}-2\cdot 2^{m})\nonumber\\
&=&A_{2^{m}}(2^{n}-2\cdot 2^{m})\nonumber\\
&=&\cdots \nonumber\\
&=&A_{2^{m}}(0).\label{eq:A6}
\end{eqnarray}
Using this result, we can derive a relation for the case of $k=2^{n}$ and $l=2^{m}+2^{m'}$ ($m'<m<n$) as follows.

\begin{eqnarray}
A_{2^{m}+2^{m'}}(2^{n})&=&\big[\bigwedge_{p=0}^{2^{m}-1}(B_{2^{m}+2^{m'}-p}\wedge C_{2^{m}+2^{m'}-p})\big]\wedge A_{2^{m'}}(2^{n}-2^{m-1})\oplus A_{2^{m}+2^{m'}}(2^{n}-2^{m})\nonumber\\
&=&\big[\bigwedge_{p=0}^{2^{m}-1}(B_{2^{m}+2^{m'}-p}\wedge C_{2^{m}+2^{m'}-p})\big]\wedge A_{2^{m'}}(0)\oplus A_{2^{m}+2^{m'}}(2^{n}-2^{m})\nonumber\\
&=&\big[\bigwedge_{p=0}^{2^{m}-1}(B_{2^{m}+2^{m'}-p}\wedge C_{2^{m}+2^{m'}-p})\big]\wedge A_{2^{m'}}(0)\nonumber\\
&&\oplus \big[\bigwedge_{p=0}^{2^{m}-1}(B_{2^{m}+2^{m'}-p}\wedge C_{2^{m}+2^{m'}-p})\big]\wedge A_{2^{m'}}(0)\oplus A_{2^{m}+2^{m'}}(2^{n}-2\cdot 2^{m})\nonumber\\
&=&A_{2^{m}+2^{m'}}(2^{n}-2\cdot 2^{m})\nonumber\\
&=&\cdots \nonumber\\
&=&A_{2^{m}+2^{m'}}(0)\label{eq:A7}.
\end{eqnarray}
Arbitrary $l<2^{n}$ is expressed as $\sum^{n-1}_{m=0} t_{m}2^{m}$ ($t_{m}=0,1$). Therefore, the relation $A_{l< 2^{n}}(2^{n})=A_{l< 2^{n}}(0)$ can be proven inductively in other cases except for Eq.~(\ref{eq:A6}) and Eq.~(\ref{eq:A7}).\par
In a similar way, the following relations for $Z_{l}(k)$ are obtained:

\begin{eqnarray}
Z_{2^{n}}(2^{n})&=&\big[\bigwedge_{p=0}^{2^{n}-1}(B_{2^{n}-p}\wedge C_{2^{n}-p})\big]\wedge A_{0}(0)\oplus B_{2^{n}}\wedge D_{2^{n}}(0)\oplus A_{2^{n}}(0)\\
Z_{l< 2^{n}}(2^{n})&=&B_{l}\wedge D_{l}(0)\oplus A_{l}(0)
\end{eqnarray}
Noting the relation $B_{l}\wedge D_{l}(2^{n})=A_{l}(2^{n})\oplus Z_{l}(2^{n})$, we derive $D_{l}(2^{n})=D_{l}(0)$.


\begin{description}
\item[Fig. 1] Quantum circuit simulating a $C^{2^{n+1}+1}$-NOT gate composed of $2^{n+2}$ MQGs. An MQG is represented as an array of $2^{n}$ $C^{2}$-NOT gates (solid or dotted frame). The number of included qubits is $2^{n+2}+1$.   
\item[Fig. 2] Example of quantum circuit simulating a $C^{5}$-NOT operation 
\item[Fig. 3] Bases of inputs and outputs in $BLOCK_{l}(k)$
\item[Fig. 4] Triangular lattice for a multi-qubits system
\end{description}


\begin{figure*}[htbp]
  \begin{center}
    \includegraphics[keepaspectratio=true,height=80mm]{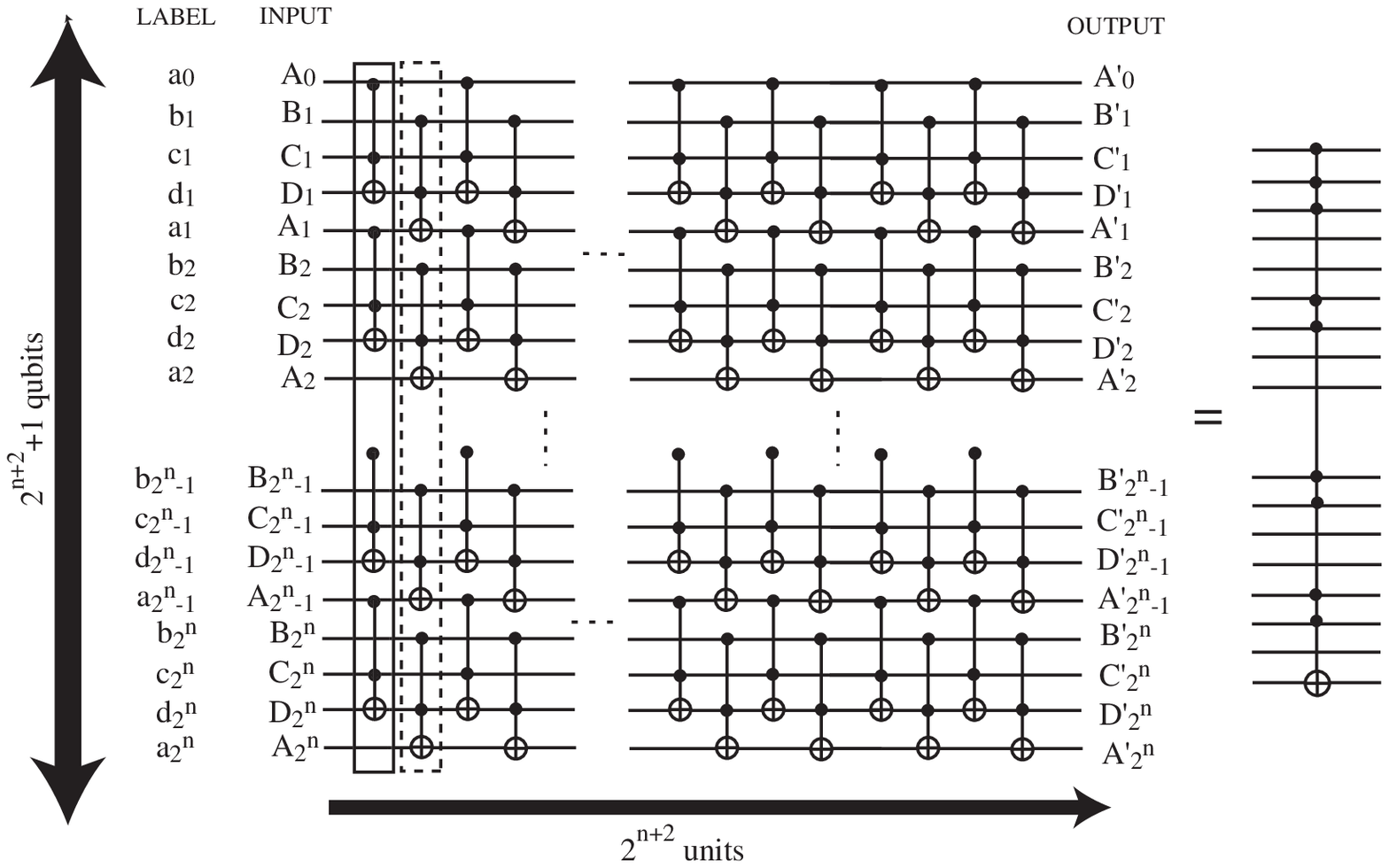}
  \end{center}
  \caption{}
  \label{fig:1}
\end{figure*}

\begin{figure}[htbp]
  \begin{center}
    \includegraphics[keepaspectratio=true,height=50mm]{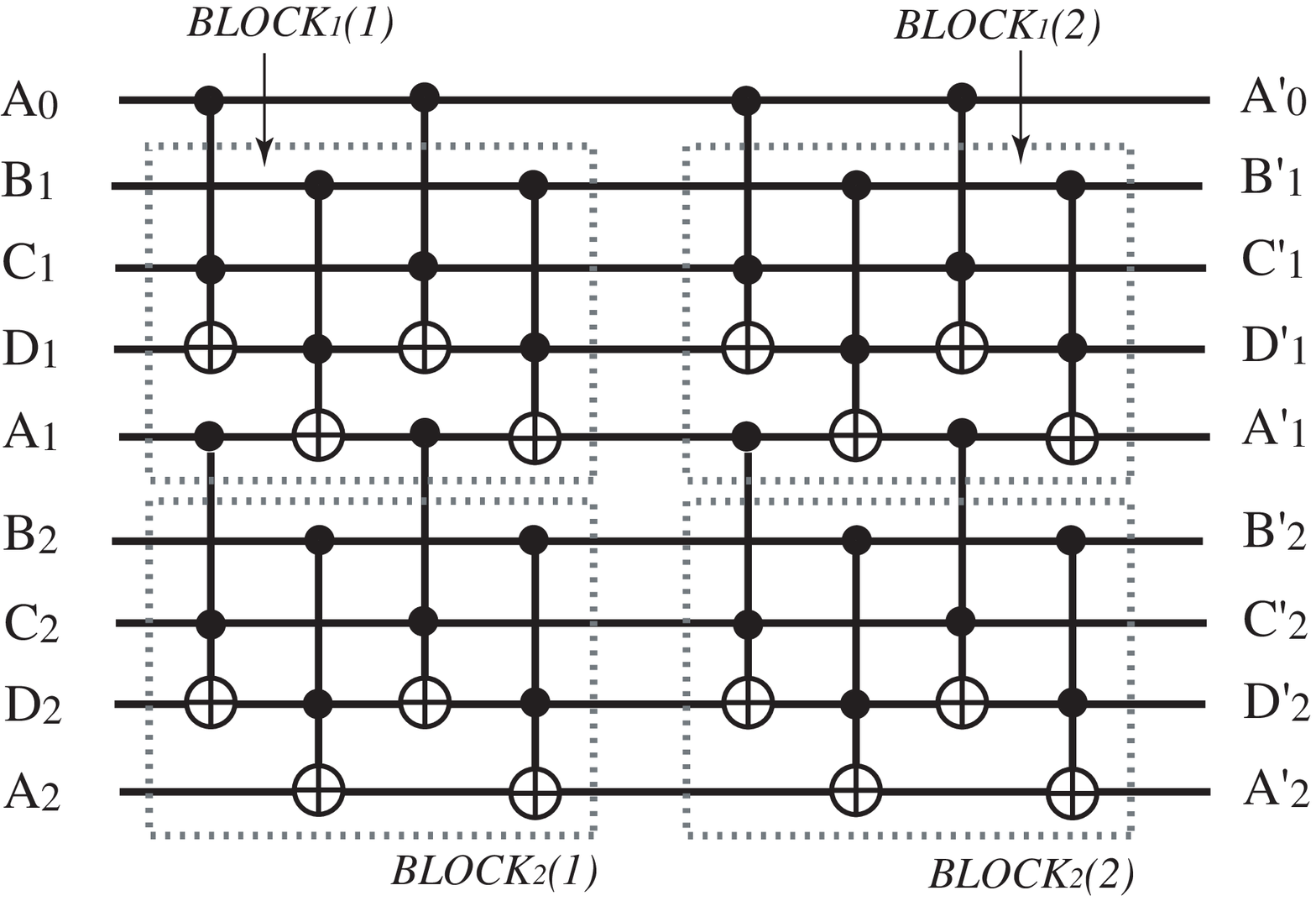}
  \end{center}
  \caption{}
  \label{fig:2}
\end{figure}

\begin{figure}[htbp]
  \begin{center}
    \includegraphics[keepaspectratio=true,height=40mm]{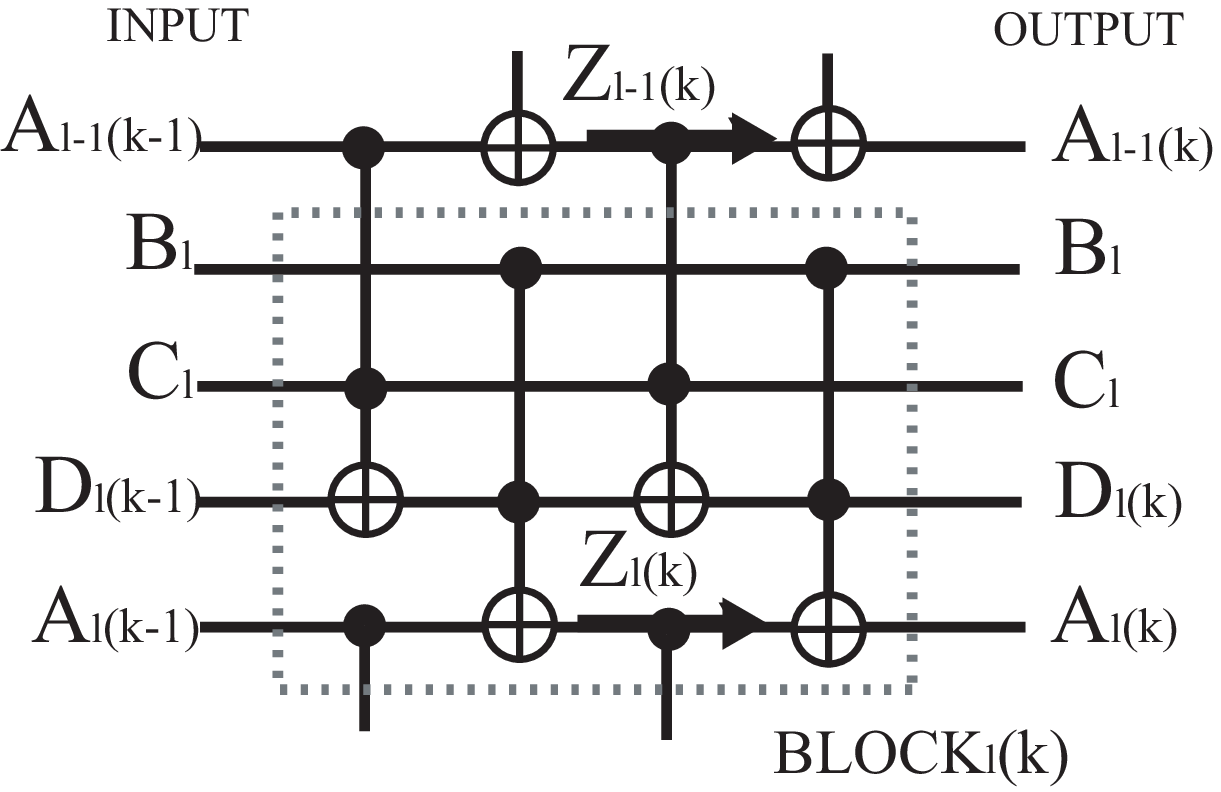}
  \end{center}
  \caption{}
    \label{fig:3}
\end{figure}

\begin{figure}[htbp]
  \begin{center}
    \includegraphics[keepaspectratio=true,height=10mm]{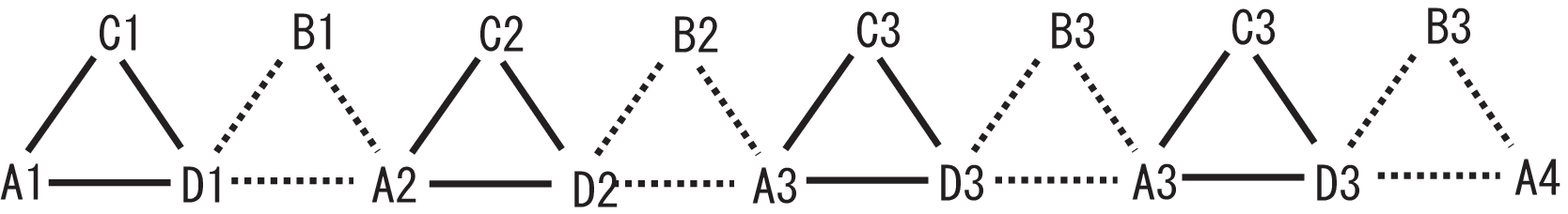}
  \end{center}
  \caption{}
  \label{fig:4}
\end{figure}

\end{document}